# Standardization of surface potential measurements of graphene domains

*Vishal Panchal[1,2], Ruth Pearce[1], Rositza Yakimova[3], Alexander Tzalenchuk[1] and Olga Kazakova[1*]*

Mr. V Panchal, Dr. R. Pearce, Prof. A. Tzalenchuk and Corresponding Author, Dr. O. Kazakova*

     V. Panchal, R. Pearce, O. Kazakova and A. Tzalenchuk
National Physical Laboratory,
Teddington, TW11 0LW, UK
E-mail: olga.kazakova@npl.co.uk
     V. Panchal
Royal Holloway, University of London,
Egham, TW20 0EX, UK
     R. Yakimova
Linköping University,
Linköping, S-581 83, Sweden

Keywords: Epitaxial graphene, Kelvin probe force microscopy (KPFM), Electrostatic force microscopy, Electrostatic force spectroscopy, Work function.

We compare the three most commonly used scanning probe techniques to obtain a reliable value of the work function in graphene domains of different thickness. The surface potential (SP) of graphene is directly measured in Hall bar geometry via a combination of electrical functional microscopy and spectroscopy techniques, which enables calibrated work function measurements of graphene domains with values $\Phi_{1LG}$ ~4.55±0.02 eV and $\Phi_{2LG}$ ~4.44±0.02eV for single- and bi-layer, respectively. We demonstrate that frequency-modulated Kelvin probe force microscopy (FM-KPFM) provides more accurate measurement of the SP than amplitude-modulated (AM)-KPFM. The discrepancy between experimental results obtained by different techniques is discussed. In addition, we use FM-KPFM for contactless measurements of the specific components of the device resistance. We show a strong non-Ohmic behavior of the electrode-graphene contact resistance and extract the graphene channel resistivity.

# 1. Introduction

Mapping of the local electronic properties of graphene is necessary for control of growth parameters and for understanding device functionality. The accurate quantification of the values measured is essential in order for the properties of graphene to be reliably understood and compared.

The growth of graphene by the sublimation of Si from SiC is arguably the most advanced method for producing continuous, homogeneous large area graphene.[1] Control over the layer thickness has been demonstrated,[2] with sublimation being the method of choice for device manufacture, where a continuous, large area single-layer (1LG) of graphene is required, *i.e.* nanoelectronics, sensing, THz applications, etc. Due to advances in the sample growth, it is generally possible to achieve homogenous single layer coverage over large areas (*i.e.* ~95% 1LG coverage for sample presented in this work). However, even small inclusions of bi-layer graphene (2LG) leads to redistribution of carriers, inhomogeneous screening effects, and the corresponding nanoscale changes in the surface potential (SP) and work function ($\Phi$). Unambiguous determination of the layer thickness of epitaxially grown graphene using atomic force microscopy (AFM) is particularly challenging due to the stepped nature of the SiC substrate coupled with growth of graphene, which often nucleates at step edges[1]. Epitaxial graphene grows due to the high temperature sublimation of Si from SiC. As three SiC layers must sublime for one layer of graphene to form, thicker layers of graphene often appear lower in topography maps, further complicating the measured height profiles of graphene layers.[3]

Scanning measurement techniques, such as Kelvin probe force microscopy (KPFM), are widely used for mapping the SP of graphene as well as identification of graphene layers. For example, KPFM has recently been used to distinguish between areas of 1LG, 2LG, few layer graphene (FLG) and the buffer or interfacial layer (0LG).[4] However, KPFM does not generally provide reliably comparable values for differences in SP between layers with a wide variety of $\Delta V_{CPD}^{1-2LG}$ values reported for 1-2LG. For example, for epitaxial graphene on SiC,

Filleter et al.[4] reported a $\Delta V_{CPD}^{1\text{-}2LG}$ = 135 meV in vacuum, whereas Burnett et al.[5] obtained a $\Delta V_{CPD}^{1\text{-}2LG}$ = 25 meV in air. On the other hand, in case of exfoliated graphene on $SiO_2$, Yu et al.[6] reported a $\Delta V_{CPD}^{1\text{-}2LG}$ = 120 meV after accounting for environmental effects by measuring in ambient atmosphere and dry nitrogen, whereas Ziegler et al.[7] reported a smaller value of $\Delta V_{CPD}^{1\text{-}2LG}$ = 68 meV in ambient conditions.

Change in the charge carrier concentration, whether it is intentional, by electrostatic or photochemical gates,[8,9] or incidental, as by uncontrolled adsorbates, modifies the measured $\Delta V_{CPD}^{1\text{-}2LG}$,[6] and the effects of substrate on the charge transfer to graphene and subsequent change of the surface potential have been discussed in depth.[7,10–12] It has also been demonstrated that atmospheric gating can modify the $\Delta V_{CPD}^{1\text{-}2LG}$ of epitaxial graphene from 0 to 100 meV on changing the environment from nitrogen to >1ppm $NO_2$ in nitrogen mixture.[13] Moreover, atmospheric humidity gating has been shown to increase the $\Delta V_{CPD}^{1\text{-}2LG}$ values.[14,15] The observed discrepancy can be attributed to differences in the used methodology and non-calibrated measurement techniques. While the reported discrepancy in the published values of $\Delta V_{CPD}^{1\text{-}2LG}$ can be partly attributed to different substrate and environmental gating, here we primarily address the accuracy of KPFM measurement technique applied to graphene domains.

We compare the ability to obtain quantified, comparable and accurate results of single-pass frequency-modulated (FM)-KPFM, conventional dual-pass amplitude-modulated (AM)-KPFM) and electrostatic force spectroscopy (EFS) by performing measurements on a graphene Hall bar device after SP calibration of the AFM probe against gold electrodes. We find that conventional AM-KPFM, being a *force* sensitive technique, suffers from a spatial averaging effect of the SP due to a significant contribution of the cantilever base and cone to the capacitive coupling, which reduces $\Delta V_{CPD}^{1\text{-}2LG}$ and leads to incorrect values of SP measured on a biased device. In contrast, FM-KPFM is sensitive to the *force gradient* and measures only the SP of the area directly under the probe apex, demonstrating improved spatial resolution and absence of averaging effects. We conclude that FM techniques, such as FM-KPFM and EFS, provide more accurate measurement of the SP than AM-KPFM. Using calibrated FM-KPFM, we perform precise work function measurements of 1LG and 2LG, being $\Phi_{1LG}$ = 4.55±0.02 eV and $\Phi_{2LG}$ = 4.44±0.02 eV, respectively. We also perform contactless measurements of the resistance of the graphene channel and two separate electrode-graphene lead contacts.

## 2. Techniques

### 2.1. Kelvin probe force microscopy

Surface potential maps of a sample can be obtained using KPFM, which measures the strength of the electrostatic forces between a conductive probe and the sample.[5] There are different methods of detecting electrostatic forces, namely: AM-KPFM, which responds to the electrostatic force at a set frequency of probe oscillation; and FM-KPFM, which responds to the electrostatic force, while maintaining constant amplitude of cantilever oscillation. As we show below, the choice of the measurement technique significantly affects the accuracy of surface potential measurements on micrometer scale graphene.

*2.1.1. Amplitude-modulated KPFM*

The AM-KPFM, discussed here, is performed as a dual-pass technique; topography line profile is mapped with tapping mode AFM during the first pass, which is then traced at a set lift height above the surface performing the surface potential measurement (**Figure 1a**). During the second pass of AM-KPFM, an AC bias voltage ($V_{AC}$ = 2 V) is applied to the probe at the mechanical resonance $f_0$ of the cantilever, causing it to oscillation due to the attractive and repulsive electrostatic interaction between the probe and the sample[16]

$$F = -\frac{1}{2}\frac{dC}{dz}\left[(V_{DC} \pm V_{CPD}) + V_{AC}\sin(\omega t)\right]^2,$$

where $V_{DC}$ is a DC bias voltage and $V_{CPD}$ is a contact potential difference between the probe and sample. A compensating $V_{DC}$ is applied to nullify the oscillations. The amplitude of the probe oscillation will be zero when the DC voltage cancels the surface electrostatic forces acting on the probe, *i.e.* $V_{DC} = V_{CPD}$. This conventional dual-pass KPFM is a well-established technique, widely used for quantitative probing of the surface potential of graphene.[1,3,4,6,17] Generally, the technique suffers from a poor lateral resolution (50-70 nm), see *e.g.* Ref. [16].

*2.1.2. Frequency-modulated KPFM*

FM-KPFM discussed here is a single-pass technique, which gives a greater degree of spatial resolution than AM-KPFM as it measures the force gradient $(dF/dz)$[18] rather than the force acting on the cantilever. Due to the geometry of the probe, the force gradient is much more confined to the probe apex, so is less affected by the parasitic capacitance of the cantilever base. The derivative of the electrostatic force decays faster with distance than the force itself and, therefore also confined to the probe apex and the sample area immediately underneath. The electrostatic forces between the probe and the sample affect the resonance frequency of the cantilever.

$$f_0 \pm f_{\text{mod}} \approx f_0\left(1 - \frac{1}{2k}\frac{dF}{dz}\right),$$

where $k$ is the spring constant of the probe. The topography is determined with the tapping mode at the main cantilever resonance, $f_0 \approx$ 70-350 kHz. Simultaneously an AC voltage with a lower frequency, $f_{\text{mod}} \approx$ 2 kHz, is applied to the cantilever with $V_{AC}$ = 8 V. This modulated frequency induces a shift in the probe resonant frequency, which appears as side lobes at frequencies $f_0 \pm f_{\text{mod}}$ (**Figure 1b**). The FM-KPFM feedback loop works to nullify these side lobes by applying an offset DC voltage to the probe ($V_{DC}$). In a similar fashion to AM-KPFM, $V_{DC}$ is recorded to generate the surface potential map. However, in contrast to AM-KPFM, FM-KPFM typically requires stiffer, higher frequency cantilevers. FM-KPFM offers a higher spatial resolution of < 20 nm as a result of force gradient localized to the probe apex and higher sensitivity to frequency shifts.[18]

AM-KPFM is usually performed as a dual-pass technique where first topography and then SP are measured along the same line in an alternating fashion, whereas FM-KPFM is most often performed as a single-pass technique, where topography and potential are recorded simultaneously, thus improving the speed of image capture. However, it should be noted that being either single- or dual-pass is not a definition of the techniques, as other examples have been demonstrated previously.[19,20]

## 2.2. EFM and EFS

### 2.2.1. Electrostatic force microscopy (EFM)

EFM is performed as a dual-pass technique: first, the topography line profile is recorded in tapping mode, and then the line profile is traced at a set lift height above the surface. During the second lifted pass, the cantilever is mechanically oscillated at $f_0$, while a constant DC bias ($V_{DC}$) is applied, probing the probe-sample electrostatic forces, which depend on the probe-sample capacitance $C$ and height $z$:[21]

$$F_{DC} = \frac{1}{2}\frac{dC}{dz}V^2, where\ V = V_{CPD} + V_{DC} + V_{induced}.$$

EFM is a purely DC technique. The electrostatic forces affect the amplitude, resonant frequency and phase of the probe. The EFM image is generated by recording the cantilever phase change[5]

$$\Delta\varphi = \frac{Q}{k}\frac{dF}{dz} = \frac{Q}{2k}\left(\frac{d^2C}{dz^2}\right)V^2,$$

where $k$ is the spring constant and $Q$ is quality factor of the cantilever. The force gradient $dF/dz$ is measured with a lock-in amplifier. Measuring the force gradient rather than the force gives sharper contrast between areas of different electrical properties, as discussed above.

However, EFS provides only qualitative information on the electronic properties of sample surface, as the individual voltage components are not separated.[18]

*2.2.1. Electrostatic force spectroscopy (EFS)*

Electrostatic force spectroscopy (EFS) is a technique that is performed at points of interest defined by EFM or other mapping techniques. Each measurement consists of oscillating the probe at $f_0$ while sweeping $V_{probe}$ and simultaneously recording $\Delta\varphi$. The plots of $\Delta\varphi$ as a function of $V_{probe}$ are parabolic, where the inflection point of the parabola is the point at which $dF/dz$ is nullified, *i.e.* the force on the probe is zero (**Figure 1c**). The inflection point is extracted post measurement and the resulting $V_{probe}$ at which $dF/dz = 0$ defines the surface potential.

EFS can be used as a quantitative and accurate measure of the SP and $\Phi_{sample}$ of a sample, if the probe is first calibrated against a sample of known $\Phi$. As EFS is not a scanning technique, probe degradation and the relevant work function change are negligible. EFS could be performed at every point of a two dimensional raster if time is not a constraint.

## 3. Results

### 3.1. AM-KPFM

**Figure 2a** shows a topography map of the graphene device. SiC step terraces are clearly visible running at a ~60° angle to the channel. Gold contacts are seen at the left and right hand sides of the image. The image reveals that it is generally rather difficult to determine the graphene layer thickness from the topography maps. Surface potential of the electrically grounded device was mapped using AM-KPFM in ambient environment (**Figure 2b**). Bright areas of 2LG are clearly visible on the 1LG background, whereas darker regions correspond to SiC trenches. The value of the $\Delta V_{CPD}^{1-2LG} = 50$ mV is consistently measured over all areas of the sample (**Figure 2c**), which is comparable to previously published results on similar samples.[10]

Further, we study the surface potential of a biased graphene device. Bias voltages of $V_{ch} = 0$, ±0.5, ±1, ±1.5 and ±2 V were applied to the left gold electrode and SP maps of the device were obtained in AM-KPFM mode. **Figure 2d** shows the plotted SP values along the marked line (Figure 2b) going through the centre of the channel and connecting the gold leads. The raw data is plotted in Figures 2c and 2d, *i.e.* no calibration of the probe's work function has been performed. As a result, the measured SP values in Figure 2d are not centred at 0 V. A significant discrepancy between applied and measured voltages is observed using AM-KPFM, *i.e.* the total difference in surface potential values measured on the left gold electrode, when biased with $V_{ch} = +2$ and $-2$ V, is only ~2.9 V, *i.e.* 27.6% less than the expected 4 V. After taking into account the work function of the probe (see Section 3.4 below), the values of $\Delta V_{CPD}$ between the biased gold contacts are still smaller than expected. This discrepancy in applied and measured voltages can be explained by the spatial averaging of AM-KPFM due to the long-range nature of the electrostatic forces acting on the probe and leading to substantial contributions of the probe cone and the base.[22] These parasitic contributions can affect the measured SP, as the area under the cantilever may not be directly over the gold leads, but instead averaging the SP over the graphene device leading to a lower total value. This spatial averaging effect is inversely proportional to the size of measured areas, *i.e.* if the area of the gold leads were large enough, so that the entire cantilever were suspended over it, a more accurate value would be expected. The same averaging effect is responsible for the SP slope on the gold leads,[22] as can be observed on both electrodes in Figure 2d.

### 3.2. FM-KPFM

Surface potential mapping has been further carried out using the FM-KPFM on the same device. **Figure 3a** shows the potential map of the grounded device. Areas of 1LG and 2LG are sharply outlined and better defined compared to the similar measurements taken with AM-KPFM. Values of $\Delta V_{CPD}^{1-2LG} = 150$ meV are recorded as shown in **Figure 3b**. This value is consistent over the device and significantly larger than $\Delta V_{CPD}{}^{1\text{-}2LG}$ obtained with AM-KPFM. The larger $\Delta V_{CPD}{}^{1\text{-}2LG}$ values can be accounted for by considering the measurement technique, which uses the force derivative rather than the force and also leads to improved spatial resolution of FM-KPFM compared to AM-KPFM. **Figure 3c** shows a line profile of the surface potential measured along the centre of the channel with $V_{ch}$ = 0, ±0.5, ±1, ±1.5 and ±2V. The change in surface potential values measured on the left gold lead when biased with $V_{ch}$ = +2 and –2 V is now ~4.18 V, *i.e.* 4.5 % larger than the expected 4 V, suggesting that this technique provides improved SP measurements even over relatively small structures with a size of several micrometres.

### 3.3 EFM and EFS

**Figure 4a** shows an EFM phase map of the grounded device. Due to the high spatial resolution, the edges of 2LG domains are sharp and well defined. **Figure 4b** shows the recorded SP values obtained from EFS measurements points over the area of 2LG and 1LG. As this is a spectroscopy rather than a mapping technique, values may be slightly affected by point position. $\Delta V_{CPD}{}^{1\text{-}2LG}$ value of 110 meV is recorded, in a good agreement with values obtained by FM-KPFM, which is expected as both techniques are sensitive to the force gradient. Results of measurements of 200 EFS spectroscopy points taken along the centre of the channel between the two gold contacts with the left contact biased at $V_{ch}$ = 0, ±1 and ±2 V are shown in **Figure 4c**.

Even the most accurate FM-KPFM and EFS techniques provide a non-zero reading of the surface potential on the grounded electrode, *i.e.* $V_{CPD}$ = –365 mV for FM KPFM (Figure 3c) and $V_{CPD}$ = –723 mV for EFS (Figure 4c). This discrepancy is the result of a work function difference between the gold and PFQNE-AL probe. Further, we account for this work function difference by subtracting from the experimental value of the $V_{CPD}$ taken at certain $V_{ch}$ the value obtained from the grounded left contact, *i.e.* $V_{CPD}(V_{ch}) – V_{CPD}(0)$. The procedure was performed using results of all three experimental techniques for the range of applied $V_{ch}$, providing $\Delta V_{CPD}$ for the left gold electrode (**Figure 5**). The measured potential drop is typically 27.6% lower than the actual $V_{ch}$ for AM-KPFM, whereas it is 4.4% and 7.8% higher for FM-KPFM and EFS, respectively. The lower $V_{CPD}$ measurements are consistent with the spatial averaging, as the relatively large base of the cantilever will weakly interact with the device channel and the right contact,[22] both of which are at a lower $V_{CPD}$ than the left contact, as it was discussed above. The higher $V_{CPD}$ measurement with FM-KPFM could be a result of an overestimation of the SP due to a relatively large excitation voltage of $V_{AC} = 8$ V and un-optimized feedback parameters, whereas the discrepancy with EFS rises from un-optimized fitting parameters.

**3.4. Work function calibration**

Further, we use FM-KPFM technique to provide accurate measurements of work function of 1LG and 2LG. Initially, work function of the PFQNE-AL probe was calibrated against the work function of the gold leads: $\Phi_{probe} \approx \Phi_{Au} + e\Delta V_{CPD}$, where $V_{CPD}$ was measured on the grounded gold electrodes. Ultraviolet photoelectron spectroscopy (UPS) measurements were carried out on four separate samples of gold deposited by e-beam evaporation under the same conditions as the deposition of the gold electrodes. The spectra were acquired with voltage of −19.04 V applied to the sample. The Fermi edge was centred at 0 eV by measuring the offset from a high resolution Fermi edge spectrum of silver calibration sample. The offset was used to correct the energy scale for all four Au spectra (**Figure 1d**). Using the indicated energies obtained from the spectra from each area of the samples, the work function was calculated. The difference in energy between the Fermi edge measured on a silver calibration sample and the cut off ($x$) is given by $x = h - \Phi$, where the energy of the incident photon is $h = 21.22$ eV. The cut off was determined by fitting a line to the relevant part of each spectrum, determining its gradient and the point at which it crosses the *x*-axis. The absolute value of *x* = 16.40 eV was obtained, thus measuring the work function of all four gold samples as $\Phi_{Au}$ = 4.82 eV. Using this value, we calculated the work function of the probe to be $\Phi_{probe}$ = 4.09 eV. Then, the work function of 1LG and 2LG were determined using the calculated values for $\Phi_{probe}$: $\Phi_{sample} \approx \Phi_{probe} - e\Delta V_{CPD}$ and the measured values of the SP extracted from the line profiles of the potential maps, *i.e.* $V_{CPD}$ = 0.20 and 0.35 V for 1LG and 2LG, respectively, see Figure 3b. This defines work functions of $\Phi_{1LG}$ ~4.55±0.02 eV and $\Phi_{2LG}$ ~4.44±0.02 eV. These work function values are within the range of previously reported results of 4.41 – 4.57 eV for 1LG measured with FM-KPFM.[6,23] However, the published values were reported to change by ±200 meV with varying lab ambient. It should be noted that the work function of graphene depends on the carrier density and exceptionally sensitive to substrate and environmental gating due to its two-dimensional nature. Thus, without reproducing substrate and environmental conditions, the work functions of graphene obtained in different experiments cannot be easily compared.

### 3.5. Contactless resistance measurements

The improved accuracy of FM-KPFM technique provides an excellent contactless method for measuring the electrode-graphene contact resistance without requiring patterned deposition of electrodes, which is typically used with the transmission line method.[6] Using experimental results shown in Figure 3c (*i.e.* line profiles of $V_{CPD}$ at $V_{ch}$ = ±2 V), contact and channel resistance can easily be deduced by normalising these line profiles [$V_{CPD}(V_{ch})$ – $V_{CPD}(0)$]/$V_{ch}$=$\Delta V/V_{ch}$ as shown in **Figure 6a**. Normalising the line profiles in this way accounts for any intrinsic $V_{CPD}$ changes, *i.e.* variations in the work function of features such as 1LG, 2LG and gold. The resulting normalised line profile is solely a consequence of the potential drop at electrode-graphene contacts and along the graphene channel due to changes in the resistivity. Dependence of the normalised voltage drop $\Delta V/V_{ch}$ as measured across the left contact, right contact and across the graphene channel (*i.e.* points 1-2, 3-4 and 2-3, respectively, in Figure 6a) are plotted in **Figure 6b**. While the voltage drop within the graphene channel is constant for all applied $V_{ch}$, this value changes linearly on electrode-graphene contacts. Careful inspection of the electrode-graphene potential drop for both contacts reveals a clear $V_{ch}$ dependence. Focusing on the left contact (points 1-2), relative change of the voltage on electrode-graphene channel is $\Delta V = 0.55$ and $-0.91$ V for $V_{ch} = +2$V and $-2$V, respectively. However, at the right contact, the $\Delta V = 0.83$ and $-0.52$ V for $V_{ch} = +2$V and $-2$V, respectively. From potential drop and $I$-$V_{ch}$ (transport) measurements (Figure 6b main panel and inset, respectively), the contactless resistance can be determined as $\Delta V/I(V_{ch})$. **Figure 6c** shows the contactless resistance measurements of the graphene channel ($R_{ch}$) only and electrode-graphene contacts ($R_{cont}$) separately for left and right contacts. The resistance of the graphene channel $R_{ch}$ ~33 kΩ is constant through the range of applied voltages, *i.e.* independent on the $V_{ch}$. The corresponding resistivity value is $\rho_{ch}$ ~ $2.7 \times 10^{-6}$ Ohm cm. On the other hand, the contacts exhibit a larger change of $\Delta R_{left\ cont}$ ~ $-17.5$ kΩ and $\Delta R_{right\ cont}$ ~17.0 kΩ as $V_{ch}$ changes from $-2$ to $+2$ V for left and right contacts, respectively. These measurements show that for this particular device, $R_{cont}$ is dependent on the $V_{ch}$ revealing a non-Ohmic behaviour. It should be noted that the edges of both gold electrodes overlap onto 2LG islands (Figure 3b), which exhibit a lower work function (Section 3.4). The lower $\Phi_{2LG}$ is a result of a high carrier density of the 2LG system, which could affect the flow of charge and, thus the contact resistance. However, it is rather difficult to give a quantitative measure of this effect from our experiments.

## 4. Conclusions

Using epitaxial graphene Hall bars with gold electrodes, we have demonstrated significant differences in accuracy and resolution between AM-KPFM and FM-KPFM techniques in determining the difference in surface potential between 1LG and 2LG. Values of $\Delta V_{CPD}^{1-2LG}$ measured with FM-KPFM demonstrate a threefold increase as compared to AM-KPFM. While AM-KPFM measures the electrostatic force on the cantilever, FM-KPFM is sensitive to the force derivative. Thus, AM-KPFM gives a weighted average of the signal, including contributions from the surface under the probe cone and cantilever. Sensitivity to the shorter range force derivative characteristic for FM-KPFM leads to spatially-confined contributions, which arise from only the probe apex, thereby reducing the spatial averaging of measured work functions observed in AM-KPFM. We experimentally demonstrate that FM-KPFM and consequently EFS have a greater degree of spatial resolution (< 20 nm) than AM-KPFM. Improvement in spatial resolution is clear on comparing the sharpness of the potential maps obtained. Moreover, we show that use of FM-KPFM and a calibrated probe provide a simple and straightforward method of obtaining an accurate measure of the work function of 1LG and 2LG. Accuracy of measurements is provided by initial calibration of the KPFM signal against the gold electrodes whose work function was measured independently by UPS. This improvement in measurement technique enables greater accuracy in calculating the work function of 1LG and 2LG, with values of $\Phi_{1LG}$ ~4.55±0.02 eV and $\Phi_{2LG}$ ~4.44±0.02eV, respectively for the device studied here. FM-KPFM was also used to investigate: i) the contact resistance between the gold electrode and graphene, revealing a non-Ohmic behavior, and ii) the resistance of the graphene channel showing Ohmic behavior with $R_{ch}$ ~33 kΩ and $\rho_{ch}$ ~ 2.7×10$^{-6}$ Ohm cm. This simple contactless method can be used to investigate the specific components of the total resistance, without fabricating devices for the transmission line method.

Our results unambiguously demonstrate that the measured values of the SP of single- and bi-layer graphene largely depend on the accuracy of the technique. However, the absolute values are also dependent on the state of the system. The SP values cannot be adequately compared unless the substrate and environmental doping is replicated. While carrying out measurements under vacuum could reduce the differences from environmental conditions, replication of precise substrate doping is challenging, which leads to a variations of the carrier concentrations and, therefore, work function values obtained in experiments.

5. Experimental Section

**Sample preparation**

Nominally monolayer epitaxial graphene was prepared by sublimation of Si and the subsequent graphene formation on the Si-terminated face of an on-axis *4H*-SiC(0001) substrate at 2000°C and 1 bar argon gas pressure. Details of the growth and structural characterization are reported elsewhere.[2] The specific synthesis route has been developed to provide large areas of homogeneous single-layer graphene. The resulting material is n-doped, owing to charge transfer from the interfacial layer,[24,25] with the measured electron concentration in the range $n = 6\text{-}20 \times 10^{11}$ cm$^{-2}$ and carrier mobility of $\mu$ ~3000 cm$^2$ V$^{-1}$s$^{-1}$ at room temperature.[26,27]

The epitaxial graphene device was fabricated by electron beam lithography, oxygen plasma etching and evaporation of Ti/Au (5/100 nm) electrodes. Details of the sample fabrication are reported elsewhere.[27] The device comprises two crosses with a channel width of 4.8 μm, surrounded by 1.6 μm-wide trench etched down into the SiC substrate.

Standard lithography fabrication methods lead to a thin (1-2 nm) layer of a resist residue on top of the graphene. The device was cleaned by sweeps the residual resist and atmospheric adsorbates from the surface of the device using contact-mode AFM prior to imaging. In order to avoid permanent damage to the device, soft contact-mode cantilevers (Bruker) with a spring constant of 0.2 N/m and a deflection set point of ~200 nm was used.

**SPM measurements**

The measurements were conducted on a Bruker Dimension Icon SPM. Doped silicon PFQNE-AL probes (Bruker) with a probe radius of ~5 nm and a spring constant of ~0.8 N/m were used for electrical measurements. Topography height images of the graphene device were recorded simultaneously with tapping phase and SP maps which were compiled from either AM-KPFM, FM-KPFM or EFM phase shift. EFS spectroscopy was conducted along the center of the device channel, *i.e.* 200 spectroscopy points were taken on the graphene channel along the marked line connecting the gold leads with the step of ~300 nm between individual points. Calibrated work function measurements of graphene were obtained with EFS by calibrating the work function of the probe against the known work function of gold electrodes, which was measured by ultraviolet photoemission spectroscopy (UPS) see Figure 1. SP measurements were carried out in ambient environment at a controlled temperature of 18 °C and humidity of ~35%.

**Transport measurements**

The transport measurements were performed in air, at room temperature, in a dark environment. The carrier density ($n$) was characterized by current biasing the device ($I_{bias}$) and measuring the Hall voltage ($V_H$) at out-of-plane magnetic fields of up to $B_{DC}$ = 0.5 T, where $n$ = $I_{bias} B_{DC}/eV_H$.[27] The carrier mobility ($\mu$) is given by $\mu = \dfrac{I_{bias}}{neV_{xx}} \times \dfrac{L}{W}$, where the longitudinal voltage ($V_{xx}$) is measured at a given $I_{bias}$. $L$ is the distance between the centers of the two crosses (20 μm) and $W$ is the width of the channel (4.8 μm).[27]

**Acknowledgements**

This work has been funded by NMS under the IRD Graphene Project (NPL) and EU FP7 Project 'ConceptGraphene'. We are very grateful to Karin Cedergren for help with the nanofabrication of graphene devices, Steve Spencer for UPS measurements and Tim L Burnett for useful discussions. We are grateful to Bruker Nano UK team for constant support of our SPM measurements.

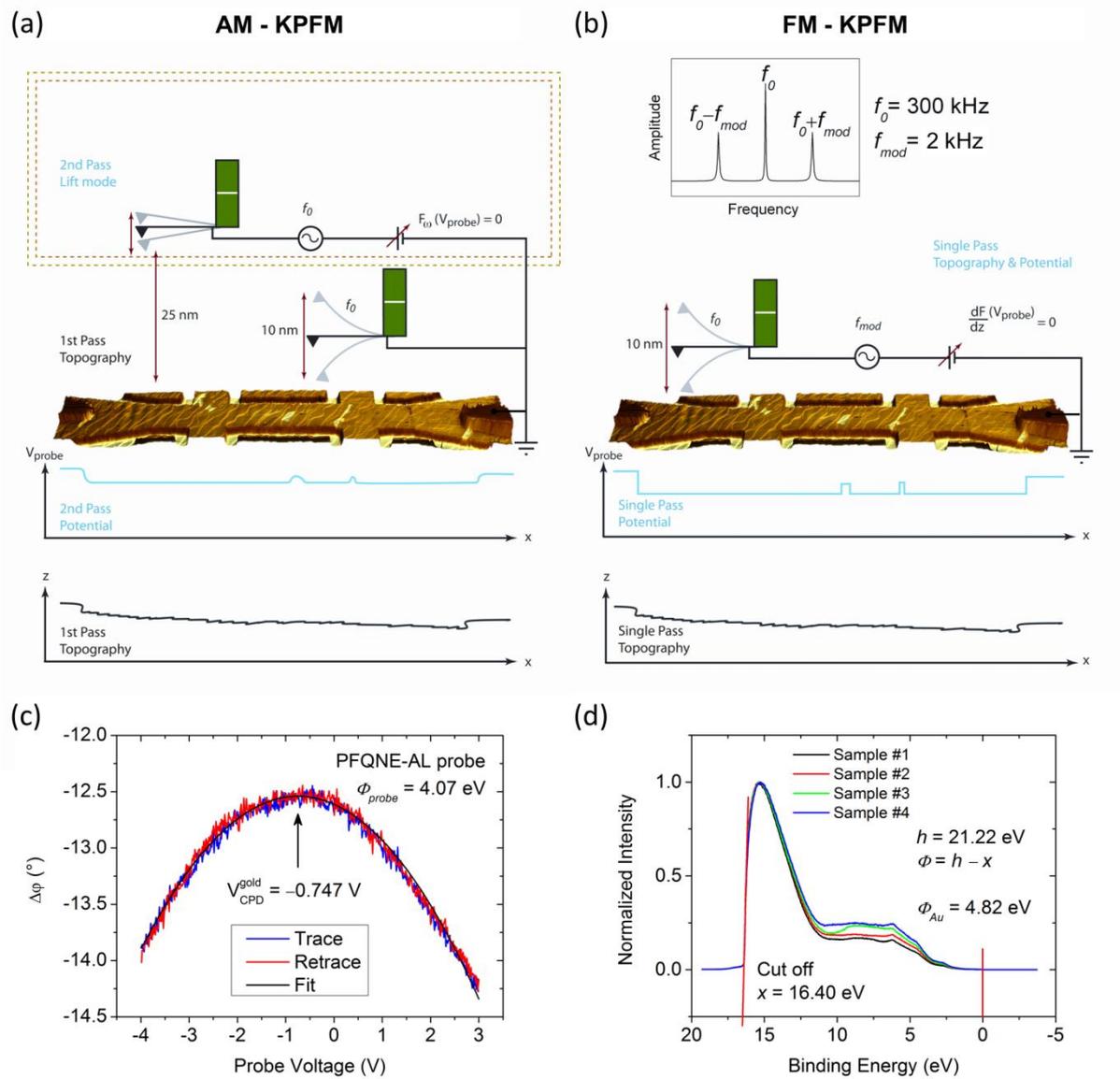

**Figure 1:** Schematic diagrams of experimental techniques: (a) AM-KPFM; (b) FM-KPFM; topography of the graphene Hall bar is superimposed with SP maps on a 3D image. Plots show characteristic profiles, *i.e.* SP on top and topography on bottom along the horizontal line in the center of the image (not shown). (c) Typical parabolic change of the cantilever phase shift measured during DC voltage sweep at a fixed point on 1LG. (d) UPS data showing the work function of gold, $\Phi_{Au} = 4.82$ eV, for four different samples.

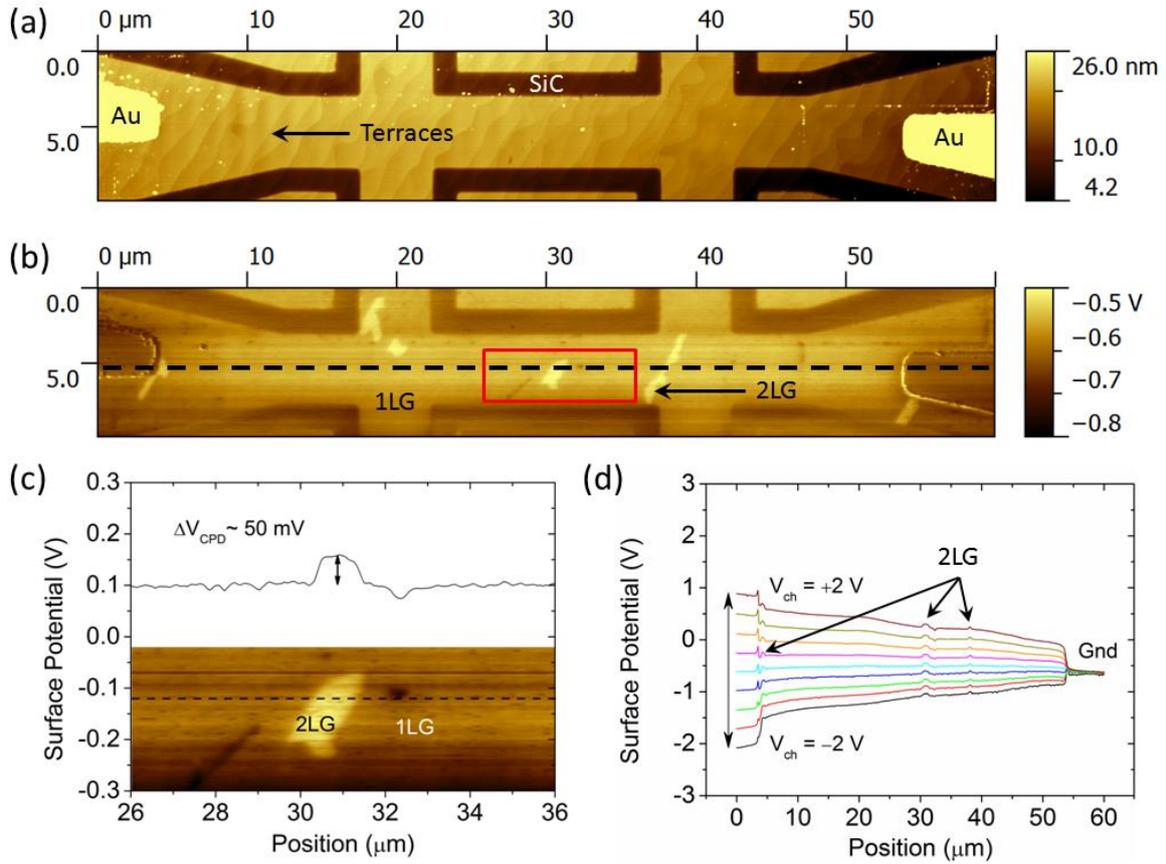

**Figure 2:** (a) Topography map of the device showing a double-cross Hall bar, gold electrodes and etched areas (bare SiC) which define the channel. (b) AM-KPFM surface potential map of the grounded Hall bar device. (c) Plot of the surface potential between areas of 1LG and 2LG within the channel along the dashed line shown in the inset. Inset shows the magnified area of the AM-KPFM surface potential map framed in (b). (d) Plot of the surface potential for the biased device measured between gold leads through the centre of the channel along the line depicted in (b), the left gold lead is biased at $V_{ch}$ between +2 and −2V and the right gold lead is grounded.

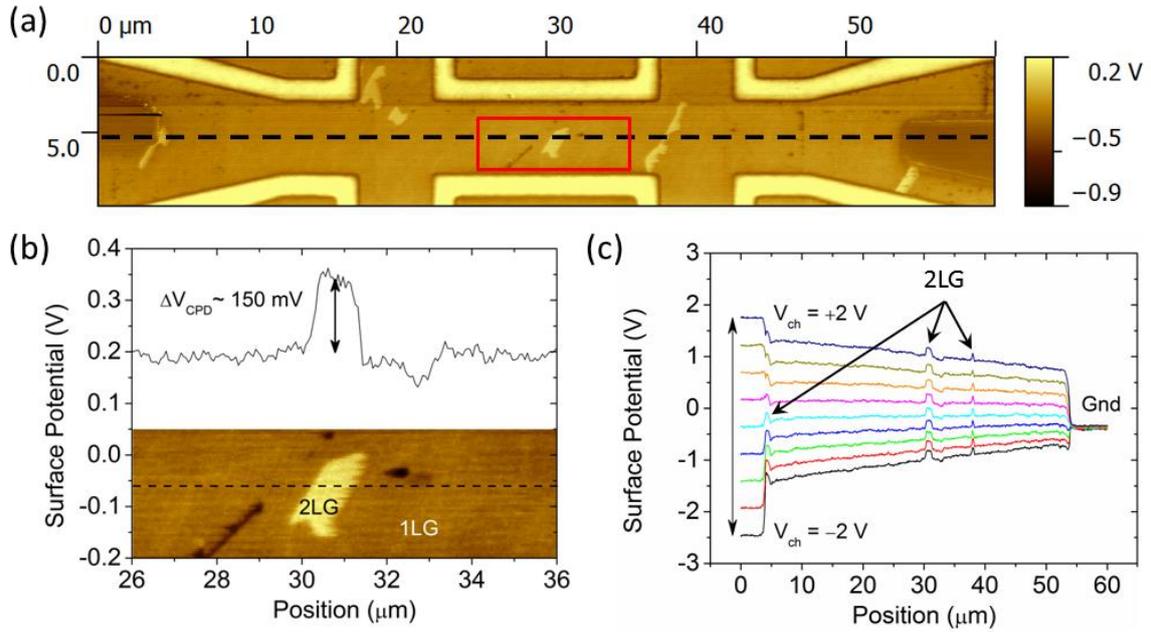

**Figure 3:** (a) FM-KPFM surface potential map of the grounded Hall bar device. (b) Plot of the surface potential between areas of 1LG and 2LG within the channel along the dashed line shown in the inset. Inset shows the magnified area of the FM-KPFM surface potential map framed in (a). (c) Plot of the surface potential measured between gold leads through the centre of the channel along the line depicted in (a), the left gold lead is biased at $V_{ch}$ between +2 and –2V and the right gold lead is grounded.

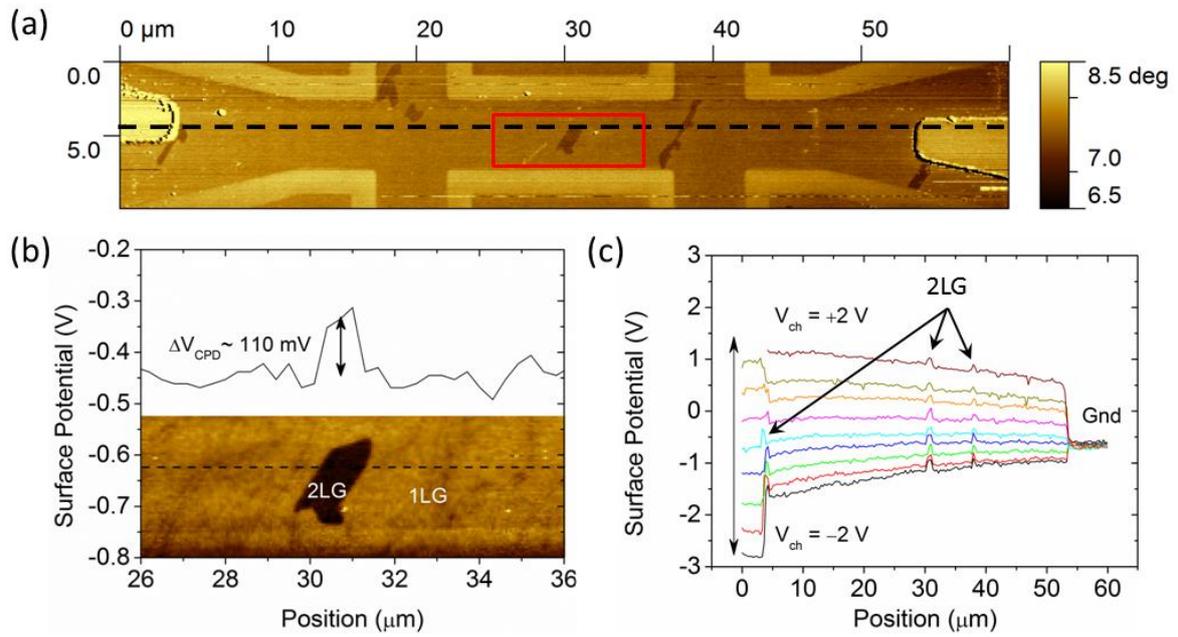

**Figure 4:** (a) EFM phase map of the grounded Hall bar device. (b) EFS plot of the surface potential between areas of 1LG and 2LG within the channel along the dashed line shown in the inset. Inset shows the magnified area of the EFM phase map framed in (a). (c) Plot of the surface potential measured by EFS between gold leads through the centre of the channel along the line depicted in (a), the left gold lead is biased at $V_{ch}$ between +2 and −2 V and the right gold lead is grounded.

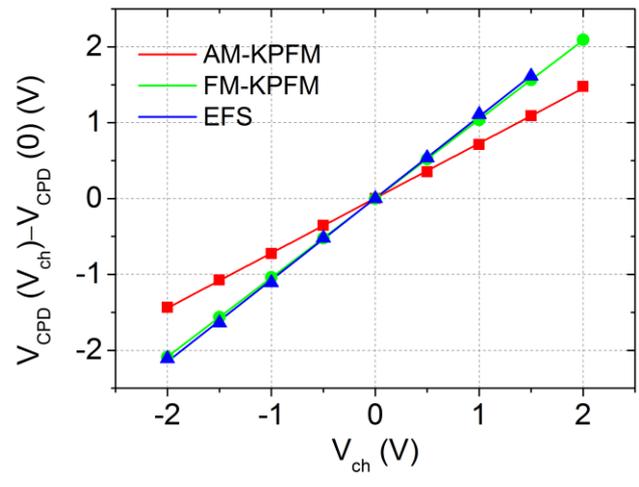

**Figure 5:** Normalised SP values as measured by AM-KPFM, FM-KPFM and EFS techniques on the left gold electrode in dependence on the voltage applied to the same gold electrode.

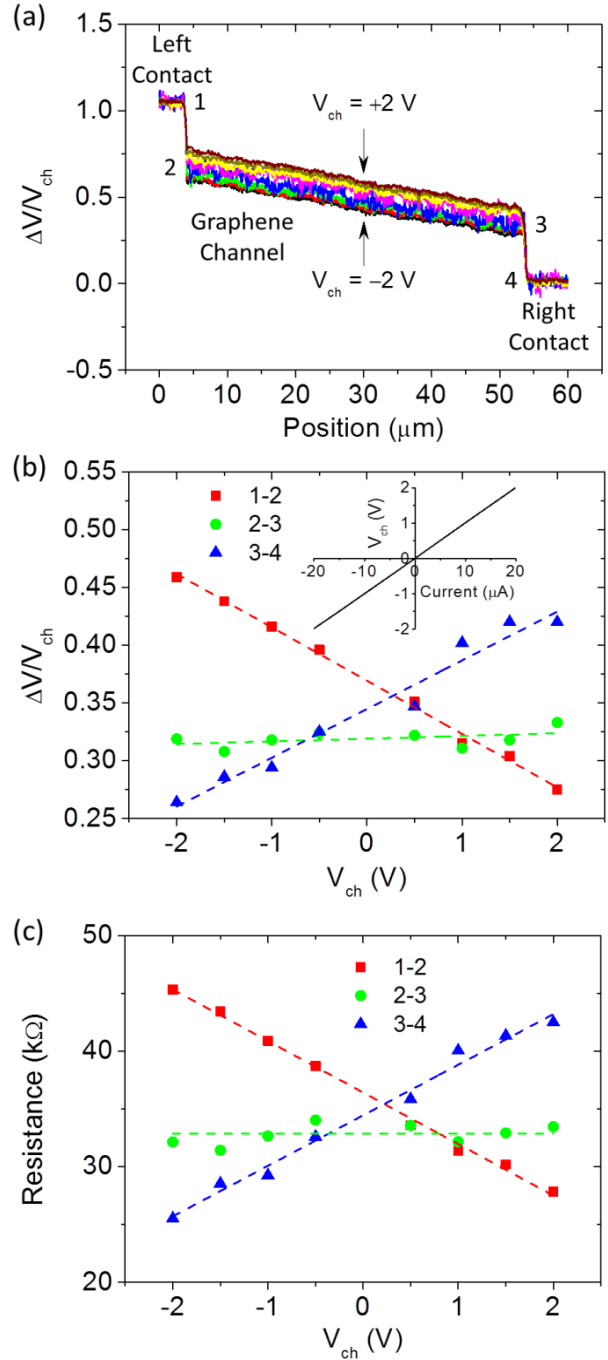

**Figure 6:** (a) Normalised surface potential line profiles. Experimental values are obtained by FM-KPFM along the dashed line in Fig. 3a. (b) $\varDelta V/V_{ch}$ and (c) resistance measurements of left contact, right contact and across the graphene channel, *i.e.* points 1-2, points 3-4 and points 2-3, respectively, in (a). Inset in (b) shows the dependence of the total current (*I*) through the circuit on the bias voltage ($V_{ch}$). Dashed lines are guides for the eye.